\documentclass[a4paper]{article}

\usepackage{amsmath,graphicx,mathtools,amssymb,setspace,lipsum,cuted,multicol,multirow,makecell,hyperref,INTERSPEECH2019,xcolor,cite}

\title{Noise Adaptive Speech Enhancement using Domain Adversarial Training}

\name{Chien-Feng Liao$^{1}$, Yu Tsao$^1$, Hung-Yi Lee$^3$, Hsin-Min Wang$^2$}

\address{
	$^1$Research Center for Information Technology Innovation, Academia Sinica, Taiwan\\
	$^2$ Institute of Information Science, Academia Sinica, Taiwan\\
	$^3$Graduate Institute of Electrical Engineering, National Taiwan University, Taiwan
	}
\email{\{r06946002,hungyilee\}@ntu.edu.tw, yu.tsao@citi.sinica.edu.tw, whm@iis.sinica.edu.tw}

\begin{document}

\maketitle
\begin{abstract}
In this study, we propose a novel noise adaptive speech enhancement (SE) system, which employs a domain adversarial training (DAT) approach to tackle the issue of a noise type mismatch between the training and testing conditions. Such a mismatch is a critical problem in deep-learning-based SE systems. A large mismatch may cause a serious performance degradation to the SE performance. Because we generally use a well-trained SE system to handle various unseen noise types, a noise type mismatch commonly occurs in real-world scenarios. The proposed noise adaptive SE system contains an encoder-decoder-based enhancement model and a domain discriminator model. During adaptation, the DAT approach encourages the encoder to produce noise-invariant features based on the information from the discriminator model and consequentially increases the robustness of the enhancement model to unseen noise types. Herein, we regard stationary noises as the source domain (with the ground truth of clean speech) and non-stationary noises as the target domain (without the ground truth). We evaluated the proposed system on TIMIT sentences. The experiment results show that the proposed noise adaptive SE system successfully provides significant improvements in PESQ (19.0\%), SSNR (39.3\%), and STOI (27.0\%) over the SE system without an adaptation.
 
\end{abstract}

\section{Introduction}
\label{sec:intro}
\begin{figure*}[t]
\includegraphics[width=0.8\linewidth,keepaspectratio]{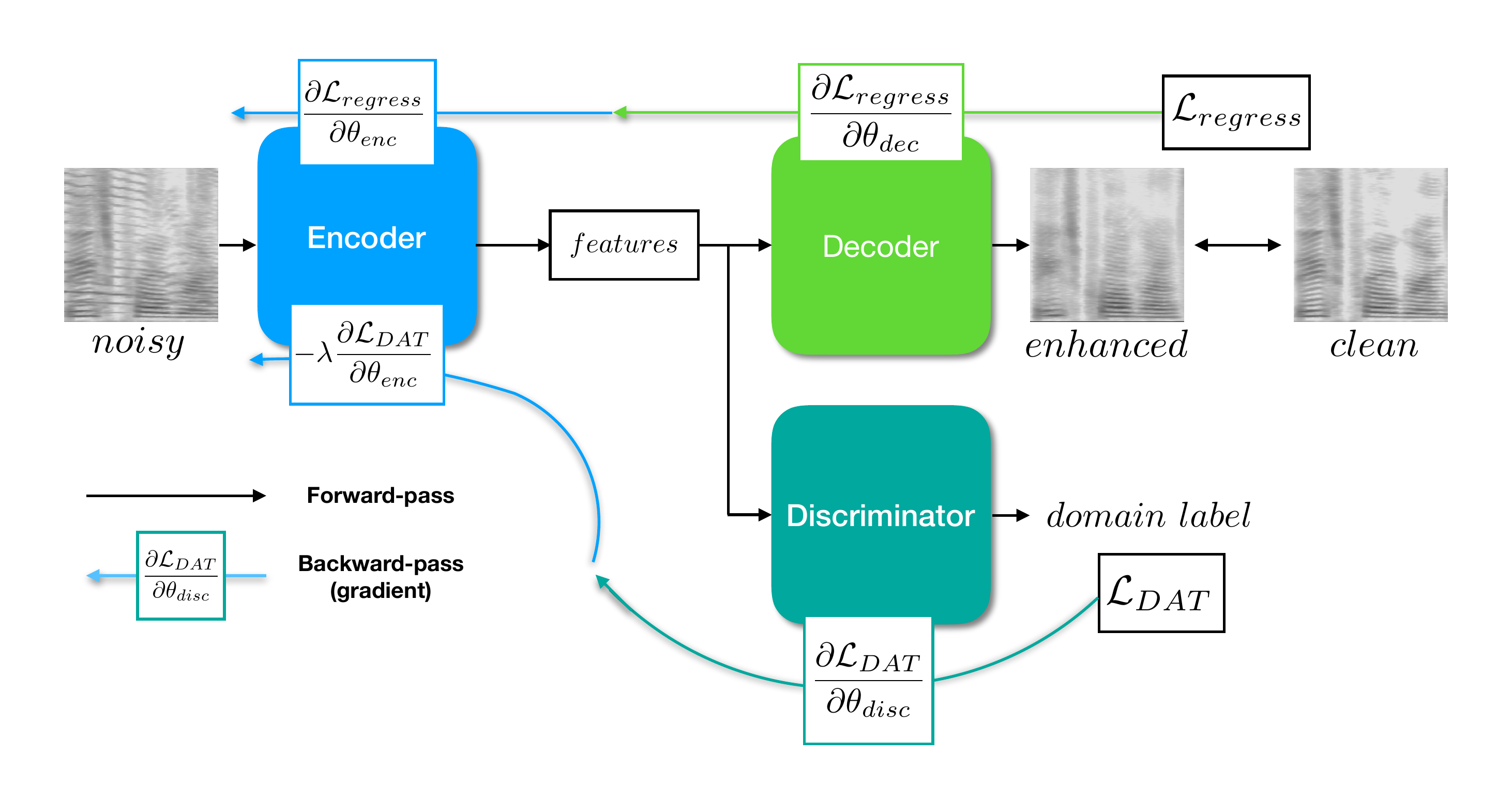}
\centering
\caption{The proposed adversarial training scheme includes an encoder, a decoder and a discriminator. During training, the encoder-decoder and the discriminator are optimized alternatively. The discriminator tries to minimize \(\mathcal{L_{DAT}}\), whereas the encoder tries to maximize it. In this way, the encoder is encouraged to produce noise-invariant features.}
\label{fig:workflow}
\end{figure*}

Speech enhancement (SE) has been widely used as a preprocessor in speech-related applications, such as speech coding, hearing aids \cite{levitt2001noise}, automatic speech recognition (ASR), and cochlear implants \cite{lai2016deep}. In the past, various SE approaches have been developed. Notable examples include spectral subtraction \cite{boll1979suppression}, minimum-mean-square-error (MMSE)-based spectral amplitude estimator \cite{ephraim1984speech}, Wiener filtering \cite{scalart1996speech}, and non-negative matrix factorization (NMF) \cite{wilson2008speech}. Recently, deep denoising autoencoder (DDAE) and deep neural network (DNN)-based SE models have also been proposed and extensively investigated \cite{lu2013speech, xu2015regression, kolbk2017speech, wang2018supervised}. However, one of the critical problems in data-driven SE is the mismatch between the training and testing environments. 

In real-world scenarios, the acoustic environment where we deploy our enhancement model can be vastly different from our training examples, and unseen noises can seriously degrade the quality of processed signal. One way to overcome this problem is to collect as many types of noises as possible to increase the generalization \cite{xu2015regression}, but it is not practical to cover potentially infinite noise types that may occur in real scenarios. We propose to come in from the domain adaptation perspective. Although not commonly seen in SE studies, this is of great interest in the field of computer vision and has been shown to be successful \cite{long2016unsupervised, tzeng2017adversarial, shu2018dirt}. The goal of domain adaptation is to utilize the unlabelled target domain data to transfer the model learned from the source domain data to a robust model in the target domain. One way is to extract domain invariant features in the use of domain adversarial training (DAT) \cite{ganin2016domain}. The key idea is to jointly train a discriminator, which can classify whether the input is from the source domain or the target domain given the extracted features, and a feature extractor, which tries to confuse the discriminator. As a result, the downstream task will not take domain information into consideration, and thus will be robust to a domain mismatch.

In our scenario, noisy utterances corrupted by stationary noises are defined as the source domain, and non-stationary noises without corresponding clean utterances are the target domain. With the help of DAT, we achieved significant improvements in objective measures including perceptual evaluation of speech quality (PESQ) \cite{recommendation2001perceptual}, segmental signal-to-noise ratio (SSNR), and short-time objective intelligibility (STOI)\cite{taal2011algorithm}. We compare our DAT with an upper-bound model and a lower-bound model. The upper-bound model is trained in a fully-supervised fashion where the clean speech references for the target domain data are available, and the lower-bound model is trained using the source domain data only, without any domain adaptation techniques. Experiments show that the proposed noise adaptive model successfully provides significant improvements over the lower-bound model.

The rest of the paper is organized as follows. We review some works focused on domain adaptation for speech-related tasks in Section \ref{sec:related}. In Section \ref{sec:domain}, we provide a detailed explanation of our approach, including the objective functions. The experiment settings and results are presented in Section \ref{sec:experiment}. Finally, we provide some concluding remarks in Section \ref{sec:conclusions}.
 
\section{Related work}
\label{sec:related}
Generative Adversarial Network (GAN) \cite{goodfellow2014generative} has recently attracted great attention in the deep learning community. Adversarial training is capable of modeling a complex data distribution by employing an alternative mini-max training scheme between a generator network and a discriminator network. One of its applications is to serve as a new objective function for a regression task. Instead of explicitly minimizing the L1$\backslash$L2 losses, which can cause over-smoothing results, the discriminator provides a high-level abstract measurement of the ``realness'' of the generated images. This idea has been used in SE tasks with parallel \cite{pascual2017segan, michelsanti2017conditional, donahue2017exploring, Meng2018AdversarialFF, pandey2018adversarial, fu2019metricgan} or nonparallel corpora \cite{mimura2017cross, Meng2018CycleConsistentSE}. 

Another important application of adversarial training is domain adaptation. In a data-driven-based pattern classification$\backslash$regression task, the mismatch between the training and testing conditions is also known as the \textit{domain shift} problem, which can cause a serious performance degradation. One way to tackle this problem is to match the data distributions across domains. Ganin et al. \cite{ganin2016domain} proposed utilizing adversarial training to produce high-dimensional features that were indistinguishable for a discriminative domain classifier. Such an idea has been deployed in various speech processing frameworks for extracting domain-invariant features. In \cite{shinohara2016adversarial, sun2017unsupervised}, the authors matched the distributions of the clean and distorted speeches in the feature space, and confirmed that the noise-invariant features were beneficial to robust acoustic models. In \cite{tsuchiya2018speaker, meng2018speaker, wang2018unsupervised, sun2018domain}, speaker-invariant and accent-invariant features were extracted in a similar fashion for speaker recognition and speech recognition. In \cite{meng2017unsupervised}, Domain Separation Network with three network components was used to extract the features. The component shared by both domains was optimized with the classification loss and DAT. To further increase the degree of domain-invariance, two private components were separately trained to be orthogonal to the shared component. Although DAT has frequently been utilized in noise-robust speech recognition, all uses have been in classification tasks. In this study, we confirmed that DAT can be deployed on an SE system as well.

\section{Domain adversarial speech enhancement}
\label{sec:domain}
We assume a scenario in which only a small amount of noisy utterances that match the testing condition (i.e., the non-stationary noise environment) are available, and there is no corresponding clean speech at all. We denote the speech dataset under stationary noises as \(S = \{x_i,y_i,c_i\}^{N}_{i=1}\), where \(x_i\) and \(y_i\) are the noisy signal and its corresponding clean signal, $c_i \in \{1,...,C\}$ indicates the noise type of the \textit{i}-th data, and $C$ denotes the number of noise classes. The unlabelled speech dataset under non-stationary noises is denoted as \(T = \{x_j, c_j\}^{M}_{j=1}\), where \(M<<N\). The goal is to utilize these unlabelled data to minimize the domain mismatch. 
\subsection{Domain adversarial training}
\label{ssec:dat}
Our network consists of three components: the encoder network \(E(x; \theta_{enc})\) whose input is noisy speech \(x\) and output is the extracted feature vector \(f\), the decoder network \(D(f; \theta_{dec})\) that estimates clean acoustic feature \(y\) given input \(f\), and the discriminator network \(Disc(f; \theta_{disc})\) that outputs a probability distribution over the domains. Here, \(\theta_{enc}\), \(\theta_{dec}\) and \(\theta_{disc}\) are the parameters of the network; for simplicity, we drop the \(\theta\) notation in the equations below. The overall workflow is shown in Figure \ref{fig:workflow}. Unlike the discriminator in \cite{ganin2016domain}, which is a binary classifier that predicts the source$\backslash$target domain, in this study it gives a probability distribution over multiple noise types. The discriminator tries to correctly predict the noise types, whereas the encoder tries to maximize the prediction error. As a result, the encoder tends to produce noise-invariant features, thereby reducing the mismatch problem.

\begin{table*}[ht]
\caption{The average PESQ, SSNR, and STOI scores for evaluating BLSTM-L, BLSTM-60, and BLSTM-U on the test set at five different SNR levels and the average scores across all SNRs. BabyCry is the adapted target domain noise type, while Cafeteria is the unseen noise type. The adaptive model (BLSTM-60) is superior to the baseline (BLSTM-L) across all SNRs for three metrics. The highest scores per metric are highlighted with bold text, excluding the BLSTM-U.}
\begin{center}
\begin{tabular}{ |c|c||  c|c|c||  c|c|c||  c|c|c|| } 
 \hline
\multicolumn{1}{|c|}{}&
    \multicolumn{1}{c||}{}&
        \multicolumn{3}{c||}{BLSTM-L (Baseline)}&
            \multicolumn{3}{c||}{BLSTM-60}&
                \multicolumn{3}{c||}{BLSTM-U (Oracle)}\\
 \cline{3-11}
& SNR(dB)	&PESQ &SSNR &STOI &PESQ &SSNR &STOI &PESQ &SSNR &STOI\\
 \hline	
\multirow{ 6}{*}{\rotatebox[origin=c]{90}{BabyCry}}
& -3    & 1.803 & -3.995 & 0.775  & \textbf{1.971} & \textbf{-0.916} & \textbf{0.802} & 2.901 & 4.981  & 0.901 \\
&  3    & 2.181 & -0.609 & 0.844  & \textbf{2.369} & \textbf{2.098 } & \textbf{0.865} & 3.208 & 6.549  & 0.929 \\
&  6    & 2.373 & 1.007  & 0.871  & \textbf{2.559} & \textbf{3.478 } & \textbf{0.890} & 3.325 & 7.195  & 0.938 \\
&  9    & 2.557 & 2.399  & 0.894  & \textbf{2.739} & \textbf{4.696 } & \textbf{0.911} & 3.419 & 7.732  & 0.945 \\
& 12    & 2.730 & 3.808  & 0.910  & \textbf{2.903} & \textbf{5.879 } & \textbf{0.925} & 3.509 & 8.314  & 0.951 \\
&Avg.   & 2.329 & 0.522  & 0.859  & \textbf{2.508} & \textbf{3.047 } & \textbf{0.879} & 3.272 & 6.954  & 0.933 \\
\hline
\multirow{ 6}{*}{\rotatebox[origin=c]{90}{Cafeteria}}
& -3    & 1.609 & -8.485 & 0.574 & \textbf{1.654} & \textbf{-7.587} & \textbf{0.603} & 1.595 & -8.703 & 0.584 \\
&  3    & 2.021 & -4.951 & 0.729 & \textbf{2.099} & \textbf{-3.595} & \textbf{0.759} & 2.031 & -4.801 & 0.745 \\
&  6    & 2.219 & -2.987 & 0.796 & \textbf{2.312} & \textbf{-1.476} & \textbf{0.820} & 2.246 & -2.574 & 0.812 \\
&  9    & 2.418 & -1.065 & 0.849 & \textbf{2.520} & \textbf{ 0.570} & \textbf{0.867} & 2.462 & -0.395 & 0.866 \\
& 12    & 2.612 & 0.715  & 0.887 & \textbf{2.720} & \textbf{ 2.476} & \textbf{0.902} & 2.669 &  1.647 & 0.905 \\
&Avg.   & 2.176 & -3.355 & 0.767 & \textbf{2.261} & \textbf{-1.922} & \textbf{0.790} & 2.201 & -2.965 & 0.782 \\

\hline
\end{tabular}
\end{center}
 \label{table1}
\end{table*}


\subsection{Objective functions}
\label{ssec:lossfunctios}
We take the regression approach as our SE objective function, which minimizes mean-absolute-error between the ground truth of the clean speech and the output of the decoder network:
\begin{equation}
\begin{split}
   \mathcal{L}_{regress} = 
   \frac{1}{N} \sum_{i=1}^{N} | D(f_i) - y_i |
   \label{eq1}
\end{split}
\end{equation}
where \(N\) is the total number of training samples from the source domain, \(x_i\) and \(y_i\) are the \(i\)-th paired noisy and clean speeches. The discriminator is optimized by minimizing the categorical cross-entropy loss, denoted as \(\mathcal{L}_{DAT}\):
\begin{equation}
\begin{split}
  \mathcal{L}_{DAT} = 
  -\frac{1}{M+N} \sum_{i=1}^{M+N} \sum_{k=1}^{C} c_{ik} log P(c_i=k | f_i)
  \label{eq2}
\end{split}
\end{equation}
where $P(c_i=k | f_i)$ is the discriminator output after softmax, and $c_{ik}$ denotes the \textit{k}-th value in one-hot expression of label $c_i$.

We attempted two methods to realize adversarial training. First, following \cite{ganin2016domain}, a gradient reversal layer (GRL) is inserted between the discriminator and the encoder. In the forward-pass, the GRL acts as an identity layer that leaves the input unchanged, but reverses the gradient passing through it by multiplying it by a negative scalar \(\lambda\) in the backward-pass. Another way is to optimize the encoder and the discriminator alternatively, as in \cite{goodfellow2014generative}. The choice of GRL instead of alternating minimization can be viewed as different approximations of mini-max \cite{fedus2017many}, and the second scheme stabilized the training and produced better results in our preliminary experiments. Finally, the network parameters are updated via gradient descent, and the overall update rules are as follows:
\begin{equation}
\begin{split}
	\theta_{enc} \leftarrow \theta_{enc} - 
	\alpha (\frac{\partial{\mathcal{L}_{regress}}}{\partial{\theta_{enc}}} 
	- \lambda \frac{\partial{\mathcal{L}_{DAT}}}{\theta_{enc}})
    \label{eq3}
\end{split}
\end{equation}
\begin{equation}
\begin{split}
	\theta_{dec} \leftarrow \theta_{dec}
	 - \alpha \frac{\partial{\mathcal{L}_{regress}}}{\theta_{dec}}
	\label{eq4}
\end{split}
\end{equation}
\begin{equation}
\begin{split}
	\theta_{disc} \leftarrow \theta_{disc} 
	- \alpha \frac{\partial{\mathcal{L}_{DAT}}}{\theta_{disc}}
    \label{eq5}
\end{split}
\end{equation}
where \(\alpha\) is the learning rate, and the hyperparameter $\lambda$ is the importance weight between two objectives.

\section{Experiments}
\label{sec:experiment}
The TIMIT database \cite{garofolo1993timit} was used to prepare the training and test sets. For the training set, 500 utterances were randomly selected from the TIMIT training set and corrupted with five stationary noise types (the source domain) at six SNR levels (-5 dB, 0 dB, 5 dB, 10 dB, 15 dB, and 20 dB) to form 15,000 source domain training data. Another 220 utterances were mixed with the non-stationary noise type (the target domain) to form the adaptation dataset. In this paper, the stationary noise types include car noise, engine noise, soft wind noise, strong wind noise, and pink noise, and the non-stationary noise type is a baby-cry noise. Hence, the total number of noise classes $C$ was six in the following experiments. For the test set, to evaluate the adaptation performance, 192 utterances from the core test set of TIMIT database were corrupted with the same non-stationary noise type (baby-cry) of the target domain. Unseen non-stationary noise type cafeteria babble was also used to evaluate the generalization ability of the adapted SE model.

\subsection{Implementation}
\label{ssec:implementation}
We used a BLSTM \cite{weninger2015speech, sun2017multiple} model as building block for the SE network. The encoder and the decoder each contains a bidirectional LSTM layer with 512 nodes, and the decoder contains a fully connected layer with 257 linear nodes at the output for spectrogram estimation. The discriminator herein was a unidirectional LSTM with 1,024 nodes, and a fully connected layer with 6 nodes followed by a softmax layer to predict the noise type. The Adam algorithm \cite{kingma2014adam} was used for training, with a learning rate of 0.0001 and 0.0005 for the SE model and the discriminator, respectively. The batch size is set to 16. We found that models are robust to the choice of $\lambda$ as long as it is below 0.1, and $\lambda$ = 0.05 was chosen through out all adaptation models.

The sampling rate of our speech data is 16 kHz. We extracted the time-frequency (T-F) features using a 512-point short time Fourier transform (STFT) with a hamming window size of 32 ms and a hop size of 16 ms, resulting in feature vectors consisting of 257-point STFT log-power spectra (LPS). For the SE model, we split each training utterance into multiple segments of 32 frames, and thus the input and output of our encoder-decoder will be a matrix of 257x32. Finally, multiple decoder outputs were concatenated and synthesized back to the waveform signal via the inverse Fourier transform and an overlap-add method. We used the phases of the noisy signals for the inverse Fourier transform. Further details and the demo may be found at \url{https://github.com/jerrygood0703/noise_adaptive_DAT_SE}. 

In the following experiments, we will evaluate our SE algorithm from three aspects: speech quality, noise reduction, and speech intelligibility. Therefore, PESQ, SSNR (in dB), and STOI will be used to evaluate the enhanced speech, respectively. For all three, a higher score indicates better results.

\begin{figure}[t]
\includegraphics[width=1.0\linewidth,keepaspectratio]{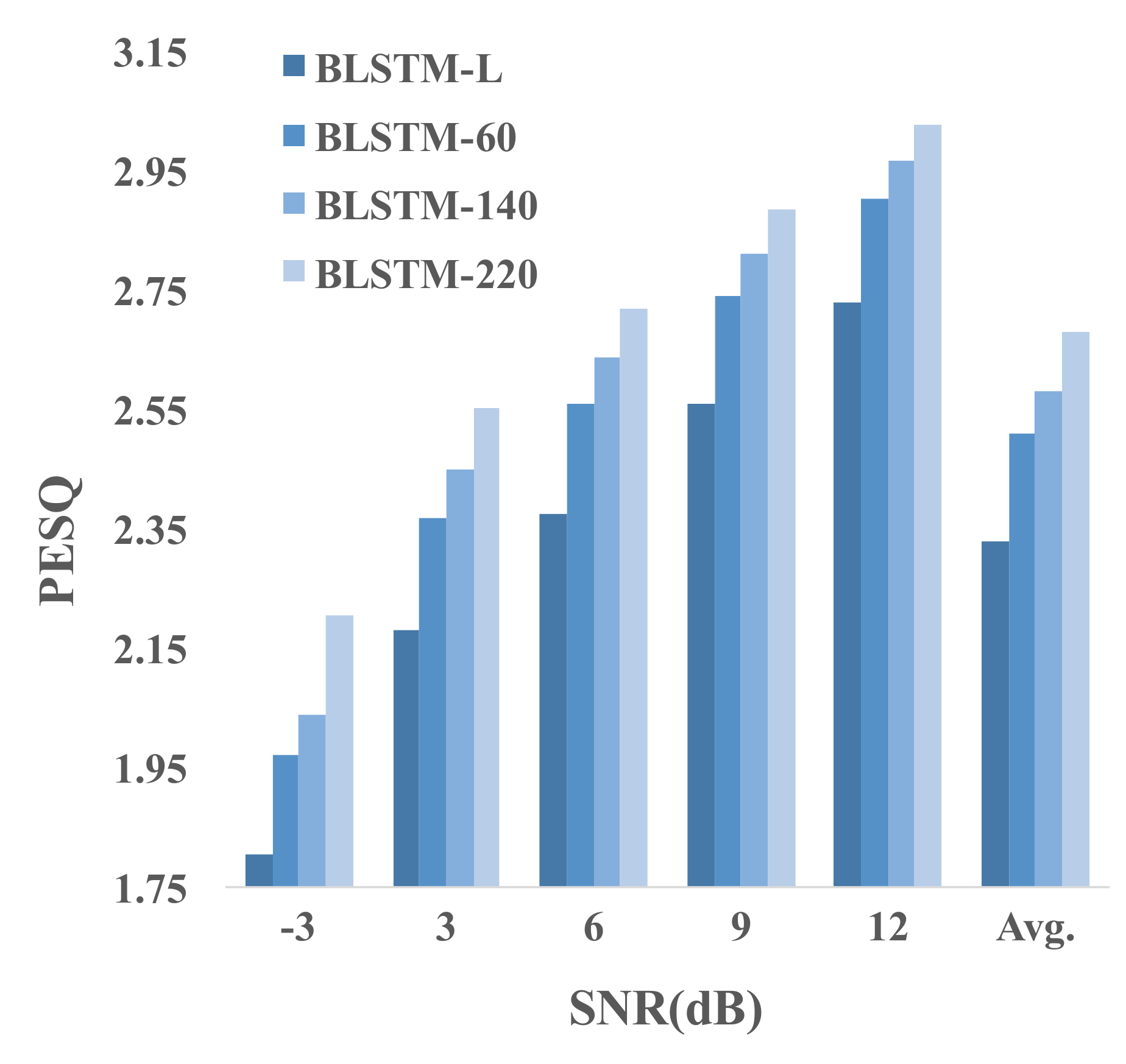}
\centering
\caption{Comparison of the baseline and the proposed models in PESQ at different SNR levels. We denote BLSTM-60 (-140 and -220) as the proposed models, where the number represents the amount of target domain noisy speech data seen during training. The PESQ scores for the unprocessed speech are 1.506, 1.882, 2.075, 2.261, and 2.456 at -3 dB, 3 dB, 6 dB, 9 dB, and 12 dB, respectively, with the average of 2.036.}
\label{fig:compare}
\end{figure}

\subsection{Results}
\label{ssec:results}

\begin{figure}[t]
\includegraphics[width=1.0\linewidth,keepaspectratio]{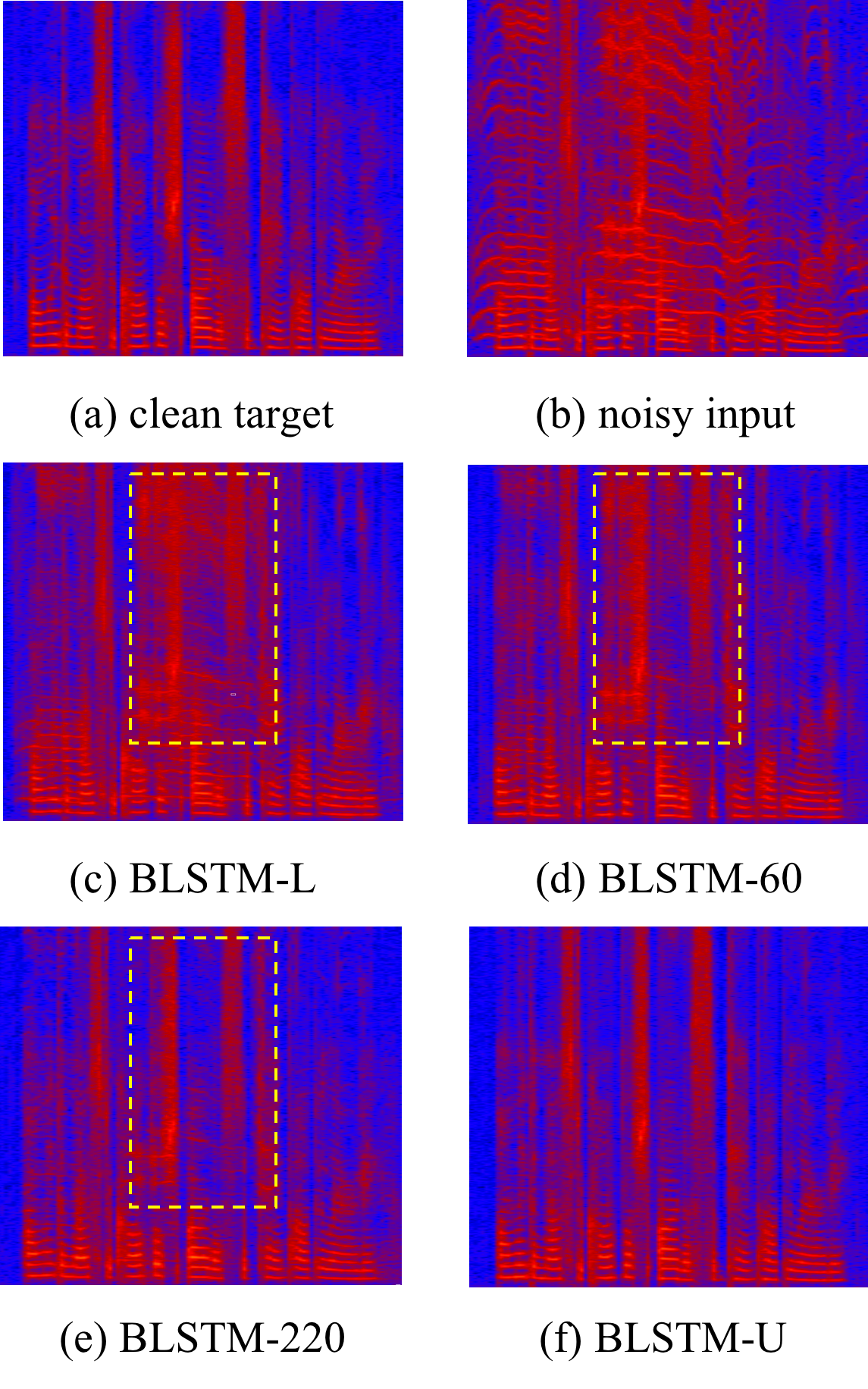}
\centering
\caption{Spectrograms of a TIMIT utterance in the teset set: (a) clean target, (b) noisy speech (baby-cry at 3 dB), (c) baseline model BLSTM-L, (d) and (e) adaptive model BLSTM-(60, 220), (f) upper-bound model BLSTM-U.}
\label{fig:spec}
\vskip -0.15in
\end{figure}

For a fair comparison, we used the same model architecture, weight initialization, and training scheme for all models compared during the experiments. The baseline is denoted as \textbf{BLSTM-L}, with $\lambda$ set to zero, meaning that only the source domain data were used to train the model parameters. First, we intend to investigate the correlation of the amount of adaptation data to the achievable performance. Thus, we prepared three sets of adaptation data. For the first set, 60 out of 220 noisy utterances were mixed with the target noise type at 0 dB SNR, which yielded 60 target domain noisy utterances that could be used in DAT. Another 80 utterances were added to formed a 140-utterances subset, while the third subset contained all of 220 utterances. The models unsupervisedly adapted with different numbers of target domain noisy utterances are dubbed as \textbf{BLSTM-60}, \textbf{BLSTM-140} and \textbf{BLSTM-220}. We also conducted a fully supervised experiment to be our upper-bound model, denoted as \textbf{BLSTM-U}. In this case, the 220 noisy training utterances of the target domain were paired with the corresponding clean utterances during training, and thus the model was optimized using fully supervised mean-absolute-error and the domain adversarial objective was not involved. 

Table \ref{table1} shows the results of the average PESQ, SSNR, and STOI scores on the test set for the \textbf{BLSTM-L}, \textbf{BLSTM-60}, and \textbf{BLSTM-U}. For the baby-cry noise set, we can see that with only 60 noisy utterances for adaptation, the proposed model outperformed the baseline at every SNR levels by a large margin. It covers 19.0\%, 39.3\%, and 27.0\% of the gap between the baseline and upper-bound model, with respect to the average PESQ, SSNR, and STOI. For the cafeteria babble noise set, we tested the SE performance using the model adapted with the baby-cry noise. The results show that the proposed model does not overfit to the adapted noise, and even attains slightly better scores. We hypothesize that DAT learns more generalized features by explicitly constraining the encoder to be noise invariant, thus making the decoder noise independent. Research into the generalization ability will be left for future studies. 

In Figure \ref{fig:compare}, we show the effectiveness of DAT on different amounts of target domain adaptation data. We can see that PESQ gradually increases when there are more target domain noisy speech data involved in training. Since unlabelled noisy speech data are easily acquired, the benefit from noise adaptive training is promising. Finally, examples of enhanced spectrograms of the baseline model, proposed methods, and upper-bound are shown in Figure \ref{fig:spec}. It is clearly shown that noises are more suppressed for the models using DAT than the baseline. In summary, the models equipped with noise adaptation achieved higher scores than the models without adaptation. 

\section{Conclusions and Future work}
\label{sec:conclusions}
In this paper, we studied the problem of noise mismatch in SE systems. We propose tackling the problem using a DAT algorithm on a BLSTM model. Utilizing unlabelled target domain noisy speech, we aimed at extracting noise-invariant features. Experimental results show that the proposed model achieves significant improvements over the baseline model using only a few noisy data. To the best of our knowledge, this is the first study adopting DAT to adapt an SE model to unseen noise types with an improved performance. In the future, we plan to study the generalization ability of SE systems with respect to unseen noise types.


\bibliographystyle{IEEEtran}
\bibliography{ref}


\end{document}